\documentclass[review,english,12pt]{elsarticle}
\usepackage[T1]{fontenc}
\usepackage[latin9]{inputenc}
\usepackage{verbatim}
\usepackage{textcomp}
\usepackage{url}
\usepackage{amsthm}
\usepackage{amsmath}
\usepackage{graphicx}

\makeatletter
\numberwithin{equation}{section}
\numberwithin{figure}{section}

\makeatother

\usepackage{babel}

\begin{document}

\title{Revisiting Multi-Subject Random Effects in fMRI: Advocating Prevalence Estimation}

\author[tau]{J.D.~Rosenblatt\corref{cor1} }
\ead{john.ros.work@gmail.com}

\author[utr]{M.~Vink}
\ead{m.vink@umcutrecht.nl}

\author[tau]{Y.~Benjamini\corref{cor1} }
\ead{ybenja@post.tau.ac.il}

\cortext[cor1]{Corresponding author}

\address[tau]{Department of Statistics and Operations Research, The Sackler Faculty of
Exact Sciences, Tel Aviv University, Israel}
\address[utr]{Rudolf Magnus Institute of Neuroscience, Department of Psychiatry, University Medical Center Utrecht, Utrecht, The Netherlands}

\begin{abstract}
Random Effects analysis has been introduced into fMRI research in
order to generalize findings from the study group to the whole population.
Generalizing findings is obviously harder than detecting activation
in the study group since in order to be significant, an activation
has to be larger than the inter-subject variability. Indeed, detected
regions are smaller when using random effect analysis versus fixed
effects. The statistical assumptions behind the classic random effects
model are that the effect in each location is normally distributed over
subjects, and ``activation'' refers to a non-null mean effect.
We argue this model is unrealistic compared to the true population variability, where, due to functional plasticity and registration anomalies, at each brain location some of the subjects are active and some are not.
We propose a finite-Gaussian--mixture--random-effect. A model that amortizes between-subject spatial  disagreement and quantifies it using the ``prevalence'' of activation at each location. 
This measure has several desirable properties: 
(a) It is more informative than the typical active/inactive paradigm. 
(b) In contrast to the hypothesis testing  approach (thus t-maps) which are trivially rejected for large sample sizes, the larger the sample size, the more informative the prevalence statistic becomes. 

In this work we present a formal definition and an estimation procedure of this prevalence. The end result of the proposed analysis is a map of the prevalence
at locations with significant activation, highlighting activations regions that are common over many brains.
\end{abstract}


\maketitle

\section{Introduction and Motivation}

A typical cognitive fMRI study entails the group-wise localization
of brain regions with evoked responses to a given cognitive stimulus.
Individual statistical maps containing regression coefficients per
voxel are combined across subjects to allow for group wise inference
using {}``Random Effects Inference''. This inference has become
standard since it offers reproducible findings (\citet{friston_classical_2002}).
Its rationale is to compare the estimated effect to its variability
over different replications with different subjects. A location is declared active if the observed
effect is improbable, compared to the sampling variability, when assuming no activation. 

The statistical assumptions behind the classic random effects approach
are: 
(a) Homogeneity: at a fixed location, all subjects are either active or inactive. 
(b) Shift alternative: {}``activation'' refers to a non-zero average effect over \emph{all} subjects.  
(c) Gaussianity: the voxel-wise effect is Gaussian distributed over subjects.

Unfortunately, these assumptions rarely hold. 
For assumption (a) to be violated, it suffice for voxels to contain different structural
and/or functional information across subjects, which is indeed the
case; As put in \citet{thirion_analysis_2007}: {}``spatial mis-registration
implies that at a given voxel, i.e. a given position in MNI space,
some subjects have activity while other subjects do not...''. Also
in \citet{fedorenko_new_2010}: {}``... activations land in similar
but mostly non overlapping anatomical locations''. The visual motion area (MT) has been noted to 
vary over individuals by more than 2 cm after Talaraich normalization. 
The primary visual cortex has also been noted to vary in size-- up to two fold over different subjects \citep{sabuncu_function-based_2009}.
Assumption (b) is just a matter of convention: should locations where only 50\% of subjects truly have non-zero activation should be called ``active'' or ``inactive''? 
Finally, regarding (c), large deviations from the Gaussianity assumption have been demonstrated
in several large studies (eg. \citet{thirion_analysis_2007} and section
\ref{sec:Results} herein). This would typically affect sensitivity but not specificity, as the t-test is known to be conservative under symmetric but heavier-than--Gaussian tailed distribution  \citep{benjamini_is_1983}. In the following, we argue this is indeed the case of group fMRI studies.

Spatial smoothing of the signal is the typical solution for the above
violations. It will spatially smear the signal so that between-subjects
agreement is larger. It will also alleviate the Gaussianity assumption
via the central limit theorem. Alas, spatial smoothing comes at the
cost of spatial precision and does not address the inherent inappropriateness of the model.

In this work we take a different path; Without spatially smoothing
the parametric maps, our model allows for voxels mapped to the same location to contain some proportion of both active and inactive individuals. The suggested model is both statistically realistic and explicitly allows for spatial disagreement over subjects-- due to either brain
plasticity or mis-registration. This amortizes the mis-registration
effects while allowing to highlight regions of agreement over subjects
with high spatial precision. We also argue that the proportion of
the active sub population at each location ({}``prevalence'') is
a more interesting parameter than the mean activation or the level of significance as appear in p-value maps. In particular when large samples are available and a significance test becomes non-informative, such as in \citet{thyreau_very_2012}. In the following sections we try to formalize and justify the population-mixture assumption.
Section \ref{sec:Method} formalizes this intuition, which is applied
to a large fMRI study in section \ref{sec:Results}. A discussion
follows in section \ref{sec:Discussion}.

\section{\label{sec:Method}Method}

\subsection{Distribution of the Voxel-Wise Effects Over Subjects}

We propose a voxel-wise adaptation of classical random-effect model.
Recalling the random effect model:\\
\begin{equation}
y_{i}(t,v)=X_{i}(t)\beta_{i}\left(v\right)+\epsilon_{i}(t,v)\label{eq:first level}
\end{equation}
Where $y_{i}(t,v)$ is the fMRI signal of subject $i$ at time $t$
in voxel $v$. The expected (unscaled) signal induced by a stimulus
is denoted by $X_{i}(t)$ and assumed known (see \citet{worsley_unified_1996}
for details). Measurement error, intra-subject psychological variability
and unmodeled effects are captured by $\epsilon_{i}\left(t,v\right)$.
The subject specific effect induced by a stimulus in voxel $v$, is
denoted by $\beta_{i}\left(v\right)$ . As previously mentioned, in
the classical random effects model it is assumed to be Gaussian distributed
over subjects, i.e. 
$\forall i=1,...,n:f(\beta_{i}) = 
\phi_{\mu,\sigma^{2}_\beta} \left( \beta_{i} \right) 
$.
% =\frac{1}{\sqrt{2\pi\sigma_{\beta}^{2}}}\exp \left( -\frac{ \left( \beta_{i}-\mu \right) ^{2}}{2\sigma_{\beta}^{2}} \right)
For the reasons described in the introduction, we will now allow it
to be a mixture of two populations: The ``inactive'' population
centred at zero and the ``active'' population with a non-zero center. 
%The model for the active population effects is as before.
After omitting the voxel index $v$, the probability density function (PDF) of this mixture is given by:

\begin{equation}
f\left(\mbox{\ensuremath{\beta}}_{i}\right)=\left(1-p\right)f_{1}\left(\beta_{i}\right)+pf_{2}\left(\beta_{i}\right)\label{eq:no-noise-pdf}
\end{equation}

Where $f_{1}\left(.\right)$ is a \emph{symmetric} distribution around zero,
$0\leq p\leq1$ 
and $f_{2}\left(.\right)=\phi_{\mu,\sigma^{2}}\left(.\right)$ with the mean effect $\mu$ allowed to be positive or negative like in the classical random effects setup.

\subsection{\label{sub:Justification}Toy Example}

\begin{figure}
\caption{\label{fig:Toy-example}Toy Signal in Two Dimensional ``Brain''.
Figure A portrays the activation region in a single arbitrary ``brain''.
Figure B portrays the true prevalence (relative frequency) of the
activation at each location after perturbing A to yield 100 ``brains'',  representing functional plasticity and misregistration. 
Figure C portrays the estimated prevalence from 100 scans with an effect variability obeying the assumptions of eq~\ref{eq:mixture PDF}:\newline
$ (1-p) \cdot (0.88 \mathcal{N}(0,0.15)+ 0.12 \mathcal{N}(0,1)) + p \cdot \mathcal{N}(1,0.25) $. \newline
Figure D is similar to C. It demonstrates the prevalence estimator's robustness to the mispecification of the active sub-population. The effect's variability is given by:\newline
$ (1-p) \cdot (0.88 \mathcal{N}(0,0.15)+ 0.12 \mathcal{N}(0,1)) + p \cdot (0.88 \mathcal{N}(1,0.15)+ 0.12 \mathcal{N}(1,1))$.
Figures E and F are the same as C and D (respectively) after masking the insignificant prevalences using Wilcoxon's signed rank statistic while controlling the FDR at 0.05 }
\centerline{\includegraphics[scale=0.6]{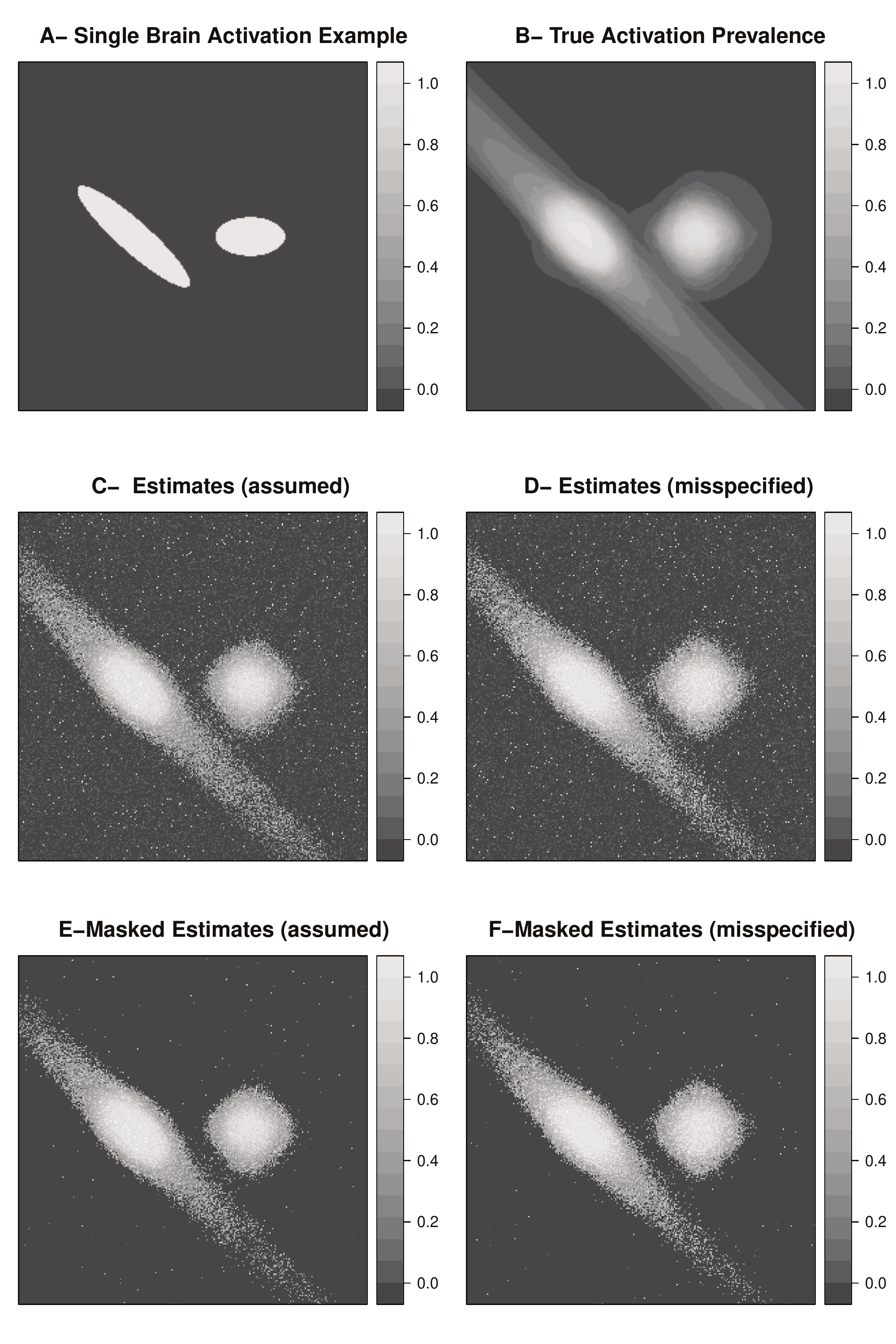}}
%\centerline{graphics[scale=0.6]{ProblemDemonstration2}}
\end{figure}

To demonstrate that eq. \ref{eq:no-noise-pdf} amortizes brain plasticity
and mis-registration consider the following example: All subjects
have a simple signal in their two dimensional ``brains'' as depicted
in figure \ref{fig:Toy-example}-A . The signal is similar across subjects, but not identical, in the
sense that different subjects are allowed to have differently ellipse-shaped
signals-- mimicking functional plasticity. The centres are also changing
slightly between subjects, mimicking misregistration effects. Figure
\ref{fig:Toy-example}-B depicts the relative frequency of the active subgroup overlap, over brains. 
We then introduce inter-subject variability for both the active and inactive subgroups and estimate $p$ using 100 ``scans''.  First, using our assumed variability model (\ref{fig:Toy-example}-C) and then using a mispecified model, to demonstrate robustness (\ref{fig:Toy-example}-D).

\subsection{Estimation Using Self-Referential Task fMRI Data}

In classical random-effects analysis, one would estimate the parameters
of interest, by deriving the marginal distribution of $y_{i}(t,v)$
mixed by $f\left(\mbox{\ensuremath{\beta}}\right)$. In the fMRI literature
it is more common to use a two-level %
\begin{comment}
convention: first level=subject. second level=population
\end{comment}
{} approach: first estimate the subject effect, and then estimate the
random effect parameters (\citet{mumford_simple_2009,xu_modeling_2009}).
These are known as the \textit{first} and \textit{second} level respectively.
We adopt this approach for convenience, both mathematical and computational, but following a referee's comment, we wish to note that proper estimation is still a matter of debate (e.g. \citet{chen_fmri_2011}) and discuss the implications of the path we chose in  \ref{apx:two-stage-estimation}.

\begin{figure}
\caption{\label{fig:KS_goodness_of_fit}Distribution of Kolmogorov Smirnov test statistic over voxels, comparing different fitted distributions to data. To avoid over fitting, a train-test data split has been used. Boxplots are sorted by the median.}
\centerline{\includegraphics[scale=0.5]{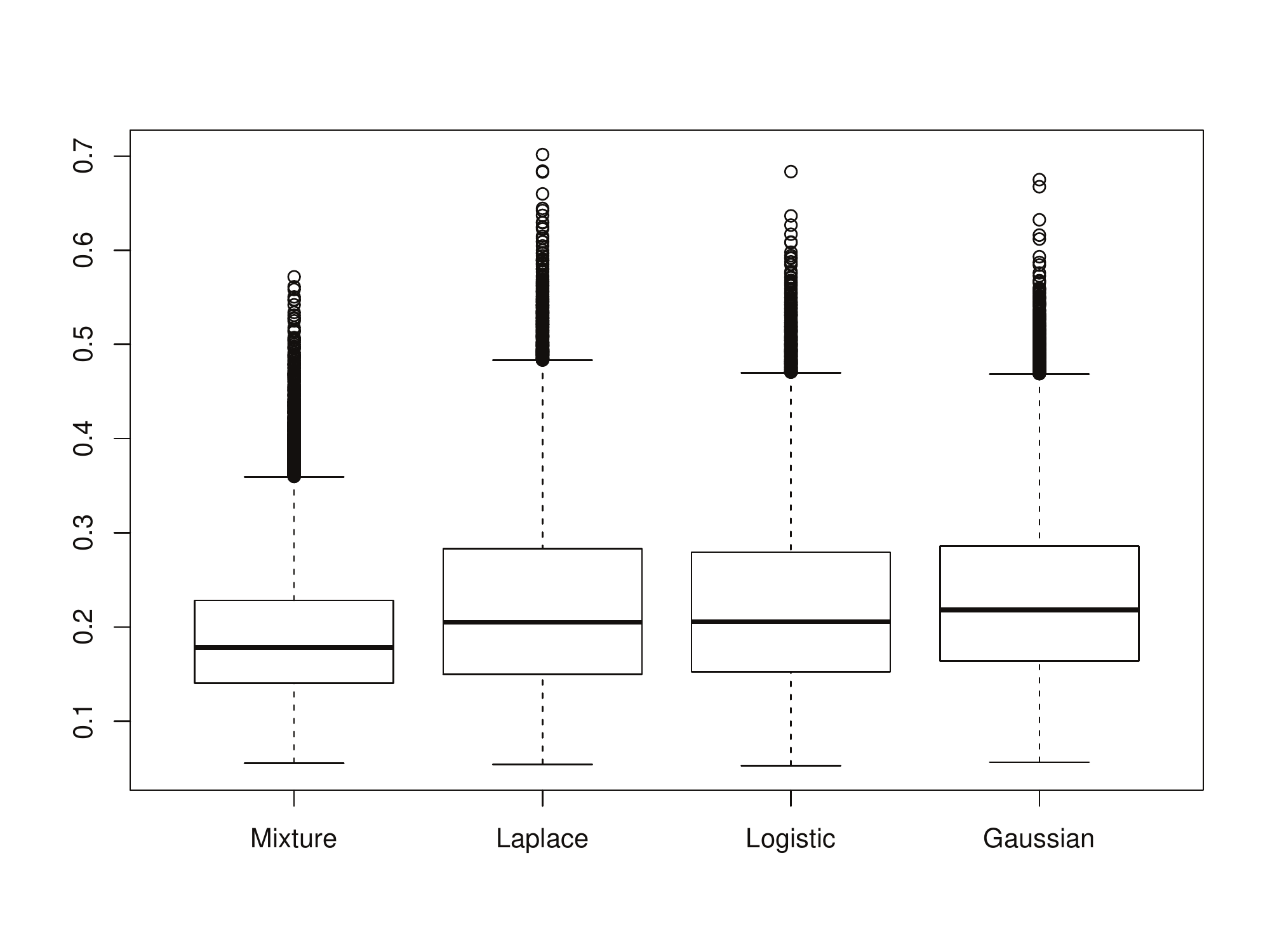}}
%\centerline{\includegraphics[scale=0.5]{CompaerdistributionKS}}
\end{figure}

Due to the two-level approach, and since $\beta_{i}\left(v\right)$ are estimated
and not observed directly, we have to allow for their measurement
error. The second level effect distribution is given by \ref{eq:no-noise-pdf}
where $f_{1}\left(.\right)$ allows for first level variance. After
considering various distributions for the inactive group-- Gaussian,
Cauchy, Laplace and Logistic-- we have chosen a centred two-component--Gaussian-scale-mixture,
which was also adopted in \citet{woolrich_robust_2008}. 
Figure \ref{fig:KS_goodness_of_fit} demonstrates the three component Gaussian mixture typically fits the data better than other candidate distributions.

By redefining $f_1(.) = \frac{1}{1-p} \left(p_1 \phi_{0,\sigma_1^2} + p_2 \phi_{0,\sigma_2^2} \right)  $,
$p_{3}=p$, and keeping the standard $f_{2}\left(.\right)=\phi_{\mu,\sigma^{2}}\left(.\right)$
we get a three component mixture:

\begin{equation}
f(\beta_i) = p_1 \phi_{0,\sigma_1^2}(\beta_i) + p_2 \phi_{0,\sigma_2^2}(\beta_i) + p_3 \phi_{\mu, \sigma_3^2}(\beta_i)
\label{eq:mixture PDF}
\end{equation}

Where $p_{1,}p_{2,}p_{3}$ are the voxel-wise mixing proportions,
naturally summing to 1, and $\phi_{mean,variance}$ are Gaussian PDFs
allowed to have voxel-specific parameters. 

We are now left with the problem of estimating the parameters of eq.
\ref{eq:mixture PDF}: $\left(p_{1},p_{2},p_{3},\mu,\sigma_{1}^{2},\sigma_{2}^{2},\sigma_{3}^{2}\right)$.
We use the expectation maximization algorithm (EM) to maximize the
likelihood. We note that as in any mixture problem, identification
problems arise. Even with the variances constrained so that $\sigma_{1}^{2}<\sigma_{2}^{2}$,
any two-component--mixture can be parametrized as $\left(p_{1},p_{2},0,\bullet,\sigma_{1}^{2},\sigma_{2}^{2},\bullet\right)\cup\left(\bullet,p_{2},\bullet,0,\sigma_{1}^{2},\sigma_{2}^{2},\sigma_{1}^{2}\right)\cup\left(p_{1},\bullet,\bullet,0,\sigma_{1}^{2},\sigma_{2}^{2},\sigma_{2}^{2}\right)$
where $\bullet$ denotes a free parameter. We alleviate this problem
by constraining the parameter space; In order to allow the interpretation
of $p_{3}$ as the prevalence of \emph{activation}, it is set to zero where the active sub-group is very similar to the inactive group. The threshing boundaries are an adaptation of the estimation limits in \citet{Donoho2004962}. Particularly $p_3=0$ if the unconstrained prevalence estimate is smaller than 
$\exp \left[-\frac{\mu^2}{2(p_1 \sigma_1^2 + p_2 \sigma_2^2)} \right] $ 
This form has the following desired qualities: 
(a) \citet{Donoho2004962}, show (figure 1) that prevalence values violating this constraint are inestimable. 
(b) It forces $p_{3}\rightarrow0$ as $\mu\rightarrow0$ permitting $p_{3}$ to be interpreted as the activation prevalence. 
(c) The constraint is more restrictive as the variance of the null population increases.

Once the prevalence has been estimated, the next natural question is {}``could it be null?''. The testing stage is a separate problem we will discuss only briefly and for which many solutions might be considered. See \citet{Roche2007501} for some examples. Note however, that unlike the classical random-effect setup, this null is not tested against a shift alternative $( H_{1}:p_{3}=1;\mu\neq0 )$,
but rather against a mixture alternative $(H_{1}:p_{3}>0;\mu\neq0 )$.
This is because we consider as {}``active'' any location with a non-null prevalence of activation. 
The generalized likelihood ratio test is not useful in this case  due to mathematical and computational complexities (see \citet{garel_recent_2007} or \citet{delmas_2003}). Instead, we use the Wilcoxon signed-rank statistic, and this for several reasons: 
(a) It is robust to model assumptions. 
(b) It is sensitive to location and \emph{shape} shifts-- both present when considering mixture alternatives. 
(c) It is easy to compute and interpret. 
(d) Surprisingly, it is more powerful than the group-t-test in our setup. We return to this point with real fMRI data in our hands in section \ref{sub:Group-T-versus-Group-Wilcoxon}.

\section{\label{sec:Results}Results}

The proposed model was used to analyze fMRI data of 64 subjects performing
a self-referential task making judgements about trait adjectives. See
 \ref{sub:Data} for details. This is an unusually large study,
offering the opportunity to validate the distributional assumptions
presented. The data has not been spatially smoothed, except for some
voxel blending due to the spatial normalization to the MNI template.
We advocate the use of unsmoothed data to avoid the notorious spatial
smearing of the signal (e.g. \citet{saxe_divide_2006}) which compromises
spatial accuracy.

\begin{figure}[h]
\caption{\label{fig:Estimated-Prevalence-in}Estimated signed-prevalence: $\hat{p_3} \cdot sign{(\hat{\mu})}$. 
Masked at significant (prevalence>0) locations using the signed-rank test statistic with FDR control
using B-H at FDR<0.1. Prevalence contour lines were added to help visualize the shape of the activation regions.}
\centerline{\includegraphics[scale=0.4]{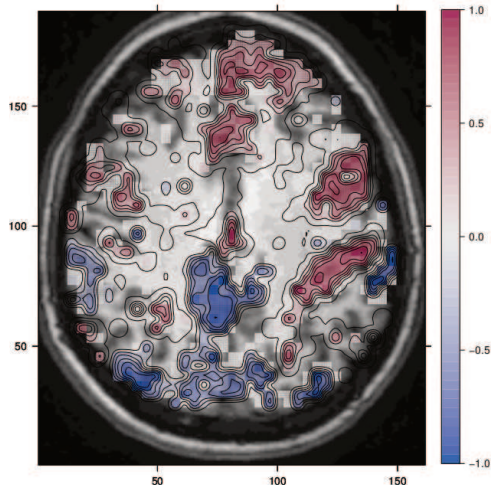}}
%\centerline{\includegraphics[scale=0.5]{prevalenceanatomyOverlay}}
\end{figure}

\begin{figure}
\caption{\label{fig:Distribution-of-Effect}Distribution of Effect in Selected
Locations: A density plot of the second-level effect distribution
(solid grey line) along with the fitted mixture (solid black line) and its two weighted inactive components (dotted lines) and a third weighted active component (dash-dotted line). The mean of the active sub population is denoted with a vertical dashed line. Group t-statistic are included.
The figures demonstrate the fit of the three component mixture to the second level effects at four locations referencing figure \ref{fig:comparison of Maps}. }
\centerline{\includegraphics[scale=0.5]{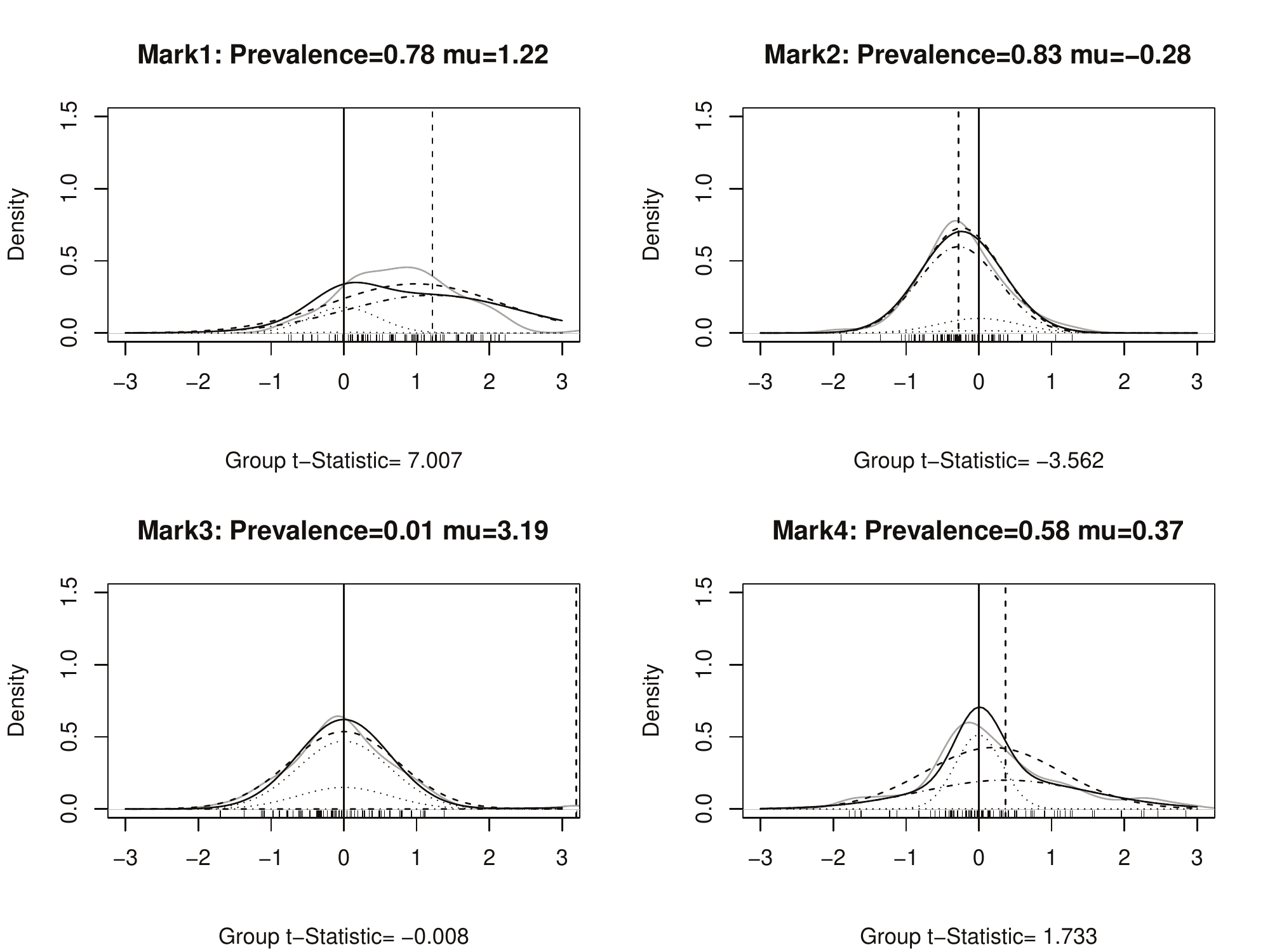}}
%\centerline{\includegraphics[scale=0.5]{effect_in_manually_marked}}
\end{figure}

The SPM of the prevalence estimates is denoted $SPM\left\{ p_{3}\right\} $
and demonstrated in figures \ref{fig:Estimated-Prevalence-in} 
and \ref{fig:comparison of Maps}-A. This estimate is compared to
the standard second-level t-statistic depicted in \ref{fig:comparison of Maps}-C.
The boundaries of the activation region exhibit a smooth decay of
$\hat{p}_{3}$ from $1$ to $0$ (more noticeable in \ref{fig:comparison of Maps}-A).
This phenomenon has already been observed by others, albeit with different
interpretation: {}``Deviation from normality of the effects... coincides
with the boundaries of activated areas'' (\citet{thirion_analysis_2007}).
Since the phenomenon is to be expected given our motivation we find
its empirical manifestation to be convincing evidence in favour of our model, where non-Gaussianity
stems from sub-populations mixing (recall, no smoothing has been applied
to the data). Also note that the change in prevalence happens at different
rates across the image which excludes voxel blending as a cause for
the smooth decay in prevalence. To further justify the mixture assumption, in figure
\ref{fig:Distribution-of-Effect} we examine the effect estimates at several select locations which indeed demonstrate the non Gaussian nature of the data. 
Figure \ref{fig:KS_goodness_of_fit} demonstrates the mixture's better fit is not limited to just some select locations, but rather occurs (on average) over the whole brain volume. We are thus confident that our mixture model is more appropriate
for the data we encounter than the single Gaussian underlying the usual random-effects analysis.

\subsection{Interpreting SPM\{prevalence\}}

\begin{figure}
\caption{\label{fig:comparison of Maps}Maps of prevalence (A) and effects
(B) compared with standard smoothed (D) and non-smoothed (C) second-level
t-maps. The distribution of contrasts over individuals and value of
t-statistic, in marked locations (1-4), can be seen in figure \ref{fig:Distribution-of-Effect}. }
\centerline{\includegraphics[scale=0.8]{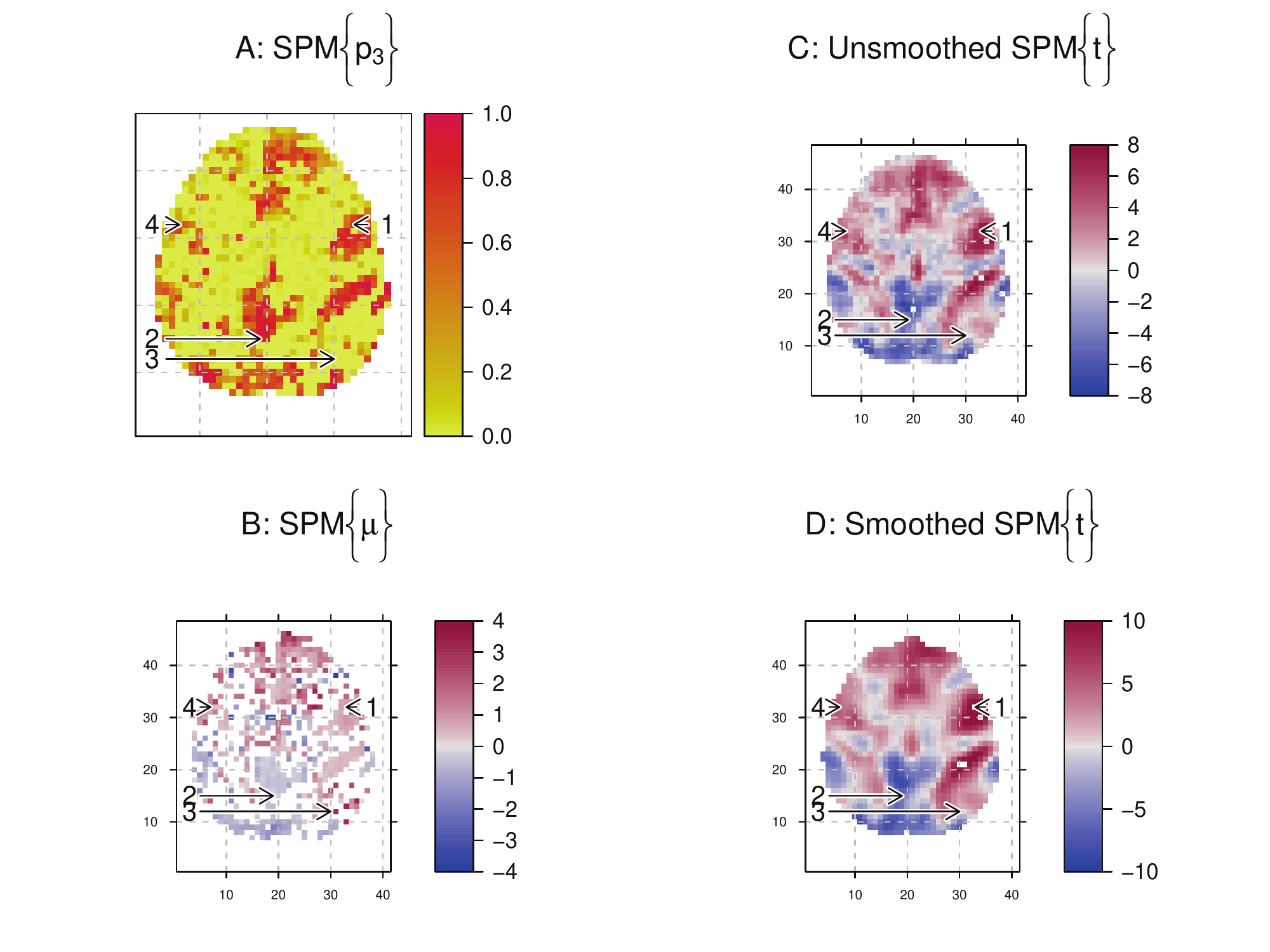}}
%\centerline{\includegraphics[scale=0.8]{compared_brain_prevalence2}}
\end{figure}

Figures \ref{fig:Estimated-Prevalence-in} and \ref{fig:comparison of Maps} depict the estimated prevalence map. A higher prevalence means more people (in the population) show
activation at that location. In particular, this says nothing about
the magnitude of the activation (when present) quantified by $\mu$. High prevalence might be accompanied by high magnitudes of effect
such as in mark 1 in fig. \ref{fig:Distribution-of-Effect} and \ref{fig:comparison of Maps}.
This is the simplest pattern of ``activation''. 
High prevalence might come with small effects (mark 2), which might be seen as statistical
artifact which will probably be weeded out by testing, or as a prevalent small effects. 
The case of large signal with small prevalence (mark 3) might be seen as an effect, say, if it were at the boundary of an activation region, or an outlier, if it were spatially isolated. 
The t-statistic generally capture the existence of signal, but note that large variability around zero, might mask
the existence of an active group such as in mark 4. 

An example of these phenomena, can be seen in figure \ref{fig:comparison of Maps}.
In particular note the activation region near coordinates $x \approx~20,y \approx~40$ (or $x\approx80,y\approx150$ in fig.~\ref{fig:Estimated-Prevalence-in}). This region is also
apparent in the t-maps in figures \ref{fig:comparison of Maps} C and D , 
albeit it is less sharp due to what is probably a small subset
of distinct (not to say {}``outliers'') subjects. Note the interesting
lobe asymmetry of this region is completely masked in the standard
smoothed $SPM\left\{ t\right\} $ in figure \ref{fig:comparison of Maps}-
D. 

Mark 2 in figure \ref{fig:Distribution-of-Effect} also demonstrates that a negative effect, or contrast, is accommodated effortlessly. A negative ``activated'' population would mean a stimulus is negatively correlated with the BOLD signal, i.e., the neurons at that voxel are inhibited. If one were interested in positive (negative) effects alone, one could consider sign, or single-sided-hypothesis masking. 

In summary, the prevalence estimate and the classical t-statistic are related, but capture different aspects of the activation pattern.

\subsection{\label{sub:Group-T-versus-Group-Wilcoxon}Group-T versus Group-Wilcoxon;
Power Considerations}

We have previously stated the Wilcoxon test should be preferred over
the group-t-test for the localization problem. To see this last point
we first note that more voxels have been found active; 11,817/27,401
using Wilcoxon versus 11,037/27,401 using the group-t-test. More importantly,
the (Pitman) asymptotic relative efficiency of the two test statistics
can be computed. We compute it using the average value of the nuisance
parameters in the current study and find it to be $e_{Wilcox,T}=0.37$.
That is, the Wilcoxon test is (asymptotically) about three times more
efficient when testing for such mixture alternatives. This result is rather surprising. We elaborate on it in a separate methodological manuscript \citep{rosenblatt_2013}.

\subsection{\label{sec:stability}Spatial Accuracy and Stability}
Visualizing the prevalence maps suggests they might display finer details than the t-maps. The boundaries of the common  regions of activation, in  figure \ref{fig:comparison of Maps}-A, seem sharper than the t-map in \ref{fig:comparison of Maps}-C (both unsmoothed). 
These details might be mere measurement noise, and thus come at the expense of stability. We would like to compare the spatial detail and stability  of the two approaches quantitatively . 

In order to compare the complexity of the emerging activation regions we computed the ratio between each region to its smallest enclosing cube. The more complicated the regions, the smaller is this ratio. We indeed find that the prevalence-defined-regions to be more complex. For instance, when half of the brain is ``active'' in the t-statistic sense, 117 out of the 286 active regions are singletons (one single voxel). This compares to 70 out of 247 such regions in the prevalence case. We also note that the median complexity of the t-regions, after excluding singletons, is 0.75 compared to a median of 0.5 for the prevalence regions. 

To establish stability we compared the agreement in the activation regions defined by the two statistics, over two split samples. We find that using half of the data (n=32) the activation regions defined by the two statistics have essentially the same stability. For instance, when half of the brain is ``active'' the agreement of the t-regions over splits is 67\% while the prevalence-regions' agreement is 60\%. We thus conclude that prevalence activation regions are indeed more complex and no-less stable.

\section{\label{sec:Discussion}Discussion}

Much of the neuroscientific literature is devoted to the localization of brain activation. Little attention is given to the magnitude of the mean effect
at active locations. This is no surprise given that the magnitude
of the effect is variable even over different sessions for the same
subject (\citet{raemaekers_test-retest_2012}). The suggested mixture-model approach
admits a natural and intuitive quantification of the extent
of activation at a location, not by its strength, but rather by its
prevalence. The typical active/inactive qualification, is an instance
of the suggested model, when $p=0$ or $p=1$. This estimation approach is particularly appropriate in large samples where prevalence estimators have low variance and significance tests are non informative and trivially rejected such as in \citet{thyreau_very_2012}. While most appealing in large studies, the stability analysis in  section \ref{sec:stability} shows that the activation regions detected using the prevalence analysis are almost as stable as the group-t regions even with about 30 subjects. Moreover, they also enjoy finer spatial detail.

The concept of {}``prevalence'' of activation is not a new one.
In \citet{friston_how_1999} the authors discuss how the use of conjunction
hypotheses could allow to infer on the population without the explicit
distributional assumptions in the random effect approach. The {}``number
of subjects in a population showing the effect'' denoted by $\gamma$
in their eq. (1), is precisely the prevalence discussed in this paper.

A test for ``at least u out of n'' active subjects was the approach taken in \citet{heller_conjunction_2007},
which could possibly be seen as a testimator of this prevalence.
Note though, that their partial conjunction inference measures the personal subject's effect against the subject's variability, while in the random effects approaches, including ours, the average effect is measured against the combined between-subject and within-subject variability.
The estimate via partial conjunction may be  therefore be 1 if all subjects' effects are significant, yet the random effect prevalence be 0 if, say, half are on the positive and the other half are symmetrically negative. The opposite may true if individual variability is large the mean effect is small yet subjects' effects are symmetrically distributed about this non-zero value. The prevalence defined in \citet{heller_conjunction_2007} is thus unrelated to our definition. For the typical definition of ``activation'', it is the latter that should be preferred.

%A test for {}``at least u out of n'' active subjects. This is indeed the approach
%taken in \citet{heller_conjunction_2007} which could possibly be seen
%as a {}``testimator'' of this prevalence. This view is however false, as conjunction inference measures the personal subject effect against the subject's variability.  While random effects measure the mean effect against both between-subject and within-subject variability. ``Conjunction prevalence'' might (truly) be 0 while our ``random effect prevalence'' is (truly) 1, and vice versa. 

The use of finite mixtures in the context of fMRI has also been suggested.
\citet{xu_modeling_2009}, motivated by the artifacts of spatial smoothing,
recur to a finite Gaussian mixture to model variability between subjects.
The number of components is however random and their weights depend
on their proximity to an {}``activation center''. The spatial distribution
of these activation centres is constructed as a multilevel point process.
This construct allows to localize both a subject's activation centres
and the group activation centres. It also admits a concept of {}``prevalence''
albeit somewhat more complicated than the one presented here.

The finite Gaussian mixture also appears in \citet{woolrich_robust_2008}.
In which the mixture is motivated by heavier-than-Gaussian tails of
the effect distribution. The author uses a scale mixture to capture
outliers in the effect distribution. The author does hint to the use
of a {}``mean-shift model'', but again, only for the purpose of
capturing outliers and not as a distinct sub population. In our work, we
have indeed adopted the Gaussian scale mixture as the null population
model, since we empirically found it to have a good fit.

\section{Acknowledgements}

We wish to thank Prof. Rafael Malach for introducing us to this problem. 

The R implementation would not have been possible without the valuable
work of Dr. Jonathan Clayden and the tractor.base package (\citet{Clayden:Maniega:Storkey:King:Bastin:Clark:2011:JSSOBK:v44i08}).

Yoav Benjamini and Jonathan Rosenblatt were supported by a European
Research Council Advanced Investigator Grant (P.S.A.R.P.S.).

\appendix
\section{\label{sub:Data}Data}

We used data from 64 subjects (30 male, 34 female, mean age 30.3 +/- 6.5 SD
years). These data were acquired at the University Medical Center
Utrecht as part of a larger study (\citet{zandbelt_reduced_2011,van_buuren_exaggerated_2011,van_buuren_reduced_2010}).
All subjects were right-handed.

Subjects performed a self-referential task. In short, subjects had
to make judgements about trait adjectives (for example \textquoteleft{}lazy\textquoteright{})
in relation to themselves (Self condition), to someone else (Other
condition), or they had to indicate whether this trait was socially
desirable (Control condition). Conditions were presented in five separate
blocks of eight trials (28s) each, alternated with rest periods of
30s. Total task duration was about 10min 32s fMRI measurements All
imaging was performed on a Philips 3.0T Achieva whole-body MRI scanner.
Functional  were obtained using a 2D-EPI-SENSE sequence with
the following parameters: voxel size 4 mm isotropic; TR= 1600 ms;
TE = 23 ms; flip angle = 72.5\textdegree{}; matrix 52x30x64; field
of view 208x120x256; 30-slice volume; SENSE-factor R=2.4 (anterior-posterior).
A total of 395 functional images were acquired during the self-reflection
task. After the acquisition of the functional images, an 3D Fast Field
Echo (FFE) T1-weighted structural image of the whole brain was made
(scan parameters: voxel size 1 mm isotropic, TR = 25 ms; TE = 2.4
ms; flip angle = 30\textdegree{}; field of view 256x150x204, 150 slices). 

fMRI preprocessing and analysis Image preprocessing and analyses were
carried out with SPM5 (http://www.fil.ion.ucl.ac.uk/spm/). After realignment,
the structural scan was co-registered to the mean functional scan.
Next, using unified segmentation the structural scan was segmented
and normalization parameters were estimated. Subsequently, all scans
were registered to a MNI T1-standard brain using these normalization
parameters and a 3D Gaussian filter (8-mm full width at half maximum)
was applied to all functional images. The preprocessed functional
images were submitted to a general linear model (GLM) regression analysis.
The design matrix contained factors modelling the onsets of the Self,
Other and Control condition as well as the instructions that were
presented during the task. These factors were convolved with a canonical
hemodynamic response function \citep{friston_1995}. To correct for
head motion, the six realignment parameters were included in the design
matrix as regressors of no interest. A high-pass filter was applied
to the data with a cut-off frequency of 0.0055 Hz to correct for drifts
in the signal. For the second-level analysis, we used the self condition
versus baseline (rest) contrast.

\section{Statistical Method}

As previously mentioned, estimation was performed by maximizing the
likelihood using an EM algorithm. A major concern when solving several
tens-of-thousands of EM problems, is speed, which is largely affected
by the initialization values. Moment estimators are typical initialization
values, but having six nonlinear moment equations these are hard to
find. We thus employ a hybrid solution, in which we search over a
grid of $\left(p_{1},p_{2}\right)$ values, solve the four moment
equations given $\left(p_{1},p_{2}\right)$, and keep the highest
likelihood value combinations as initialization values for the EM.
This initialization heuristic allowed considerable speed gains during
estimation.

Implementation was done in the R programming environment (\citet{r_development_core_team_r:_2011}).
The described estimation procedure has been implemented in the R package
\textit{FPF (fMRI Prevalence Finder)} available from R-Forge (\citet{RJournal:Theussl+Zeileis:2009})
at \url{http://rosenblatt1.r-forge.r-project.org/}. See the package's
in-line help for details. The results in this paper were obtained using version 0.53 of the package. The raw data used is included in the package for reproducibility.

\section{\label{apx:two-stage-estimation}Implications of two stage estimation}

As presented in section~\ref{sec:Method}, we use the first level effect estimates to fit a population distribution and estimate the prevalence. In particular, we do not use the first level variance estimates as done in some software suits (\citep[see][footnote 1]{worsley_general_2002}). While chosen due to its simplicity, there are several considerations supporting our approach. 
First, there are the classical considerations: (a) the between-subject variance assumed to be of larger magnitude than the within-run variance and (b) the matter affecting only efficiency and not bias. 

More importantly- the effect of inverse variance weighting is unclear, as we are no longer in the variance-components setup and we are no longer interested in the effect ($\mu$), but rather in the prevalence ($p$).

\bibliographystyle{model2-names}
\bibliography{VinkPaper3}

\end{document}